\documentclass[12pt, authoryear]{elsarticle}

\makeatletter
\def\ps@pprintTitle{%
 \let\@oddhead\@empty
 \let\@evenhead\@empty
 \def\@oddfoot{\centerline{\thepage}}%
 \let\@evenfoot\@oddfoot}
\makeatother

\usepackage{amsmath}
\usepackage{graphicx}
\usepackage{verbatim}
\usepackage{subfigure}
\usepackage{color}
\usepackage{setspace}
\usepackage{float}
\usepackage{natbib}
\usepackage{footnote}
\usepackage{hanging}
\usepackage{enumitem}
\usepackage{siunitx}
\usepackage{hyperref}
\usepackage{caption}
\makesavenoteenv{tabular}
\makesavenoteenv{table}

\usepackage[top=2.3cm,left=2.3cm,right=2.3cm,bottom=2.3cm]{geometry}

\begin{document}

\begin{frontmatter}

\title{Evaluation of short-term temporal evolution of Pluto's surface composition from 2014-2017 with APO/TripleSpec}

\author[stsci]{B. J. Holler}
\author[usc]{M. D. Yanez}
\author[swri]{S. Protopapa}
\author[swri]{L. A. Young}
\author[uva]{A. J. Verbiscer}
\author[nmsu]{N. J. Chanover}
\author[low]{W. M. Grundy}

\address[stsci]{Space Telescope Science Institute, Baltimore, MD, USA}
\address[usc]{University of Southern California, Los Angeles, CA, USA}
\address[swri]{Southwest Research Institute, Boulder, CO, USA}
\address[uva]{University of Virginia, Charlottesville, VA, USA}
\address[nmsu]{New Mexico State University, Las Cruces, NM, USA}
\address[low]{Lowell Observatory, Flagstaff, AZ, USA}

\begin{abstract}
\doublespacing{In this work we present the results of a spectral observing campaign of Pluto to search for temporal changes in surface composition on 1- to 3-year timescales. Near-infrared spectra of Pluto were obtained from June 2014 to August 2017 with the TripleSpec cross-dispersed spectrograph at the Apache Point Observatory's 3.5-meter Astrophysical Research Consortium (ARC) telescope. Observations were requested in order to obtain spectra of approximately the same sub-observer hemisphere $\sim$14 months apart, thus removing the effects of viewing geometry and rotation phase. Comparison of the CH$_4$ (methane) band areas and band center shifts between each component of these ``matched pairs" revealed a surface in transition. Band areas for the 1.66 and 1.72 $\mu$m CH$_4$ absorption features exhibited a $>$5-$\sigma$ increase between 2014-06-17 and 2015-08-19, corresponding to a sub-observer hemisphere centered at $\sim$280$^{\circ}$E, with the latter date only 1 month after the New Horizons flyby of Pluto. The majority of matched pairs were obtained of the anti-Charon hemisphere, home to the bright, volatile-rich Sputnik Planitia, and did not present statistically significant changes in CH$_4$ band areas. CH$_4$ band center shifts, which provide information on the mixing state of CH$_4$ and N$_2$ in solid solution, were calculated between components of each matched pair, with no significant band shifts detected. The favored explanation for these combined results is the sublimation of more-volatile N$_2$ from the northern latitudes of Pluto in the lead-up to northern hemisphere summer solstice in 2029, leading to an increase in CH$_4$ concentration.}
\end{abstract}

\begin{keyword}
Pluto, surface; Ices, IR spectroscopy; Kuiper belt; Trans-neptunian objects
\end{keyword}

\end{frontmatter}

\section{Introduction}
\begin{figure}[ht!]
\begin{center}
\includegraphics[scale=0.5,trim=0cm 0cm 0cm 0cm,clip=true]{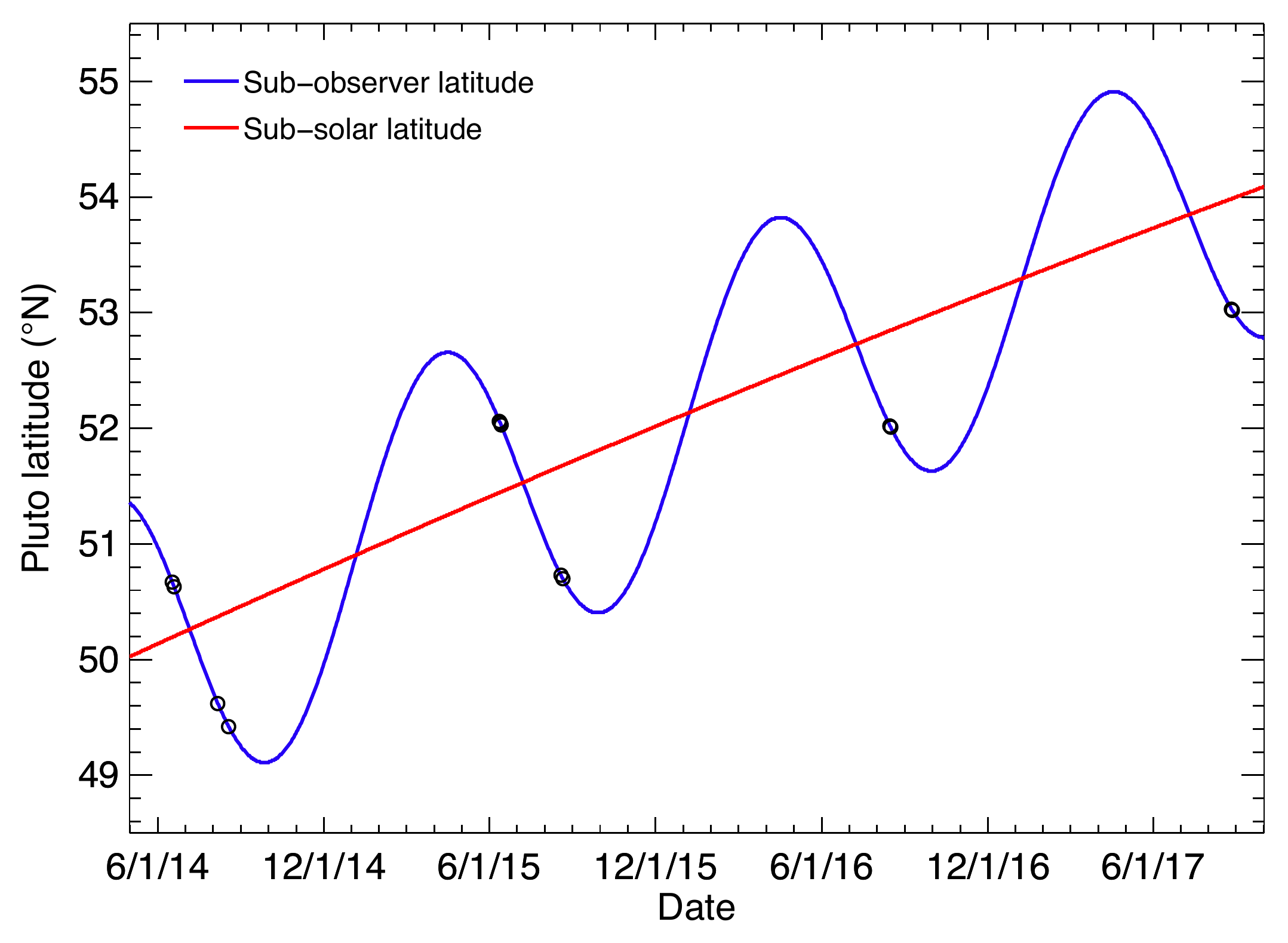}
\caption{Comparison of the sub-solar (red) and sub-observer (blue) latitudes on Pluto as a function of date. The date range covers June 2014 to August 2017, with open black circles marking the times of observations discussed in this work. The sub-solar latitude is monotonically increasing during this period, while the sub-observer latitude exhibits sinusoidal variations of $\pm$1.5$^{\circ}$ due to parallax produced by Earth's orbital motion. (Values retrieved from JPL Horizons.)\label{subsolobs}}
\end{center}
\end{figure}

\doublespacing{Pluto was first identified over 90 years ago and since that time has completed less than forty percent of its full orbit (248 Earth-years). Studies of Pluto's surface composition have been ongoing for only a small subset of this period, beginning with the first evidence for CH$_4$ (methane) ice on the surface (Cruikshank et al., 1976). This was only a few years before Pluto reached equinox in 1988 and perihelion in 1989, so the majority of spectroscopic studies have taken place during northern hemisphere spring, as Pluto recedes from the Sun and the sub-solar latitude migrates northward. Since Pluto has a $\sim$122$^{\circ}$ obliquity, and its heliocentric orbit is eccentric ($e$=0.25), these seasonal transitions ought to be extreme (e.g., Stern and Trafton, 1984; Binzel et al., 2017) and potentially observable over time. Models of Pluto's surface and atmospheric evolution suggest that volatile ices (N$_2$, CO, and CH$_4$) should be lost from the northern hemisphere, except Sputnik Planitia, as Pluto approaches northern hemisphere summer solstice in 2029 (e.g., Hansen and Paige, 1996; Young, 2013; Hansen et al., 2015; Bertrand and Forget, 2016). Given the detection of extensive deposits of volatile ices in the northern hemisphere by New Horizons in 2015 (Grundy et al., 2016; Protopapa et al., 2017), the removal process must occur relatively rapidly, if the models are correct. It is worth noting that the duration of the New Horizons flyby was too short to observe large-scale changes in surface composition or ice distribution. Visible imaging data were used to search for extremely short-lived phenomena, such as geysers and cryovolcanic eruptions, but no instances of these events were positively identified (Hofgartner et al., 2018).\\
\indent Grundy et al. (2014) analyzed ground-based near-infrared spectra of Pluto over a 13-year period and presented evidence of a temporal decrease in the strength of the N$_2$ and CO ice bands, as well as a decrease in the blueshift of the CH$_4$ ice bands, possibly indicating an increase in CH$_4$ concentration in CH$_4$:N$_2$ mixtures (e.g., Quirico and Schmitt, 1997; Protopapa et al., 2015). The rate of change also appeared to be accelerating between 2010 and 2014, when the last spectra analyzed in the paper were obtained. While changes on the order of 5+ years are evident, Grundy et al. (2014) note that the effects of Pluto's rotational variability may have been imperfectly accounted for, preventing evaluation of changes on shorter (yearly) timescales. Longer-term evaluation of temporal changes over 30+ years from 1983 to 2014 was reported by Lorenzi et al. (2016) using visible spectra of Pluto. Their results indicate a general reddening of Pluto's surface over time and a decrease in band depth for the weak CH$_4$ features in this region of the spectrum. However, the significant changes in sub-observer latitude between 1983 ($\sim$10$^{\circ}$S) and 2014 ($\sim$50$^{\circ}$N) more than likely influenced their CH$_4$ band depth comparisons.\\
\indent One method for evaluating short-term changes while taking into account rotational variability and changes in viewing geometry is to obtain spectra of Pluto at the same sub-observer latitude and longitude roughly a year apart, as first demonstrated by Grundy and Buie (2001) and later by Grundy et al. (2013). This is made possible due to the parallax produced by the Earth's motion in its orbit about the Sun, with the resulting shifts in viewing angle nearly parallel to Pluto's spin axis, as projected on the sky. At the present time this leads to a limited range of sub-observer latitudes ($<$3$^{\circ}$) on Pluto repeating approximately 14 months later (Figure~\ref{subsolobs}). Grundy and Buie (2001) reported 4 such sets of ``matched pairs" between 1997 and 1998 while Grundy et al. (2013) presented 3 sets between 2005 and 2006, but neither reported a statistically significant detection of temporal changes. In order to quantify Pluto's short-term surface changes, while correcting for Pluto's rotational variability, we designed a spectroscopic observing program specifically to make use of these matched pairs over the course of several years (2014-2017). These data provide a valuable temporal context to the observations made by the New Horizons spacecraft during its flyby of Pluto in July 2015 (e.g., Stern et al., 2015).}

\section{Observations}
\begin{figure}[ht!]
\begin{center}
\includegraphics[scale=0.65,trim=0cm 0cm 0cm 0cm,clip=true]{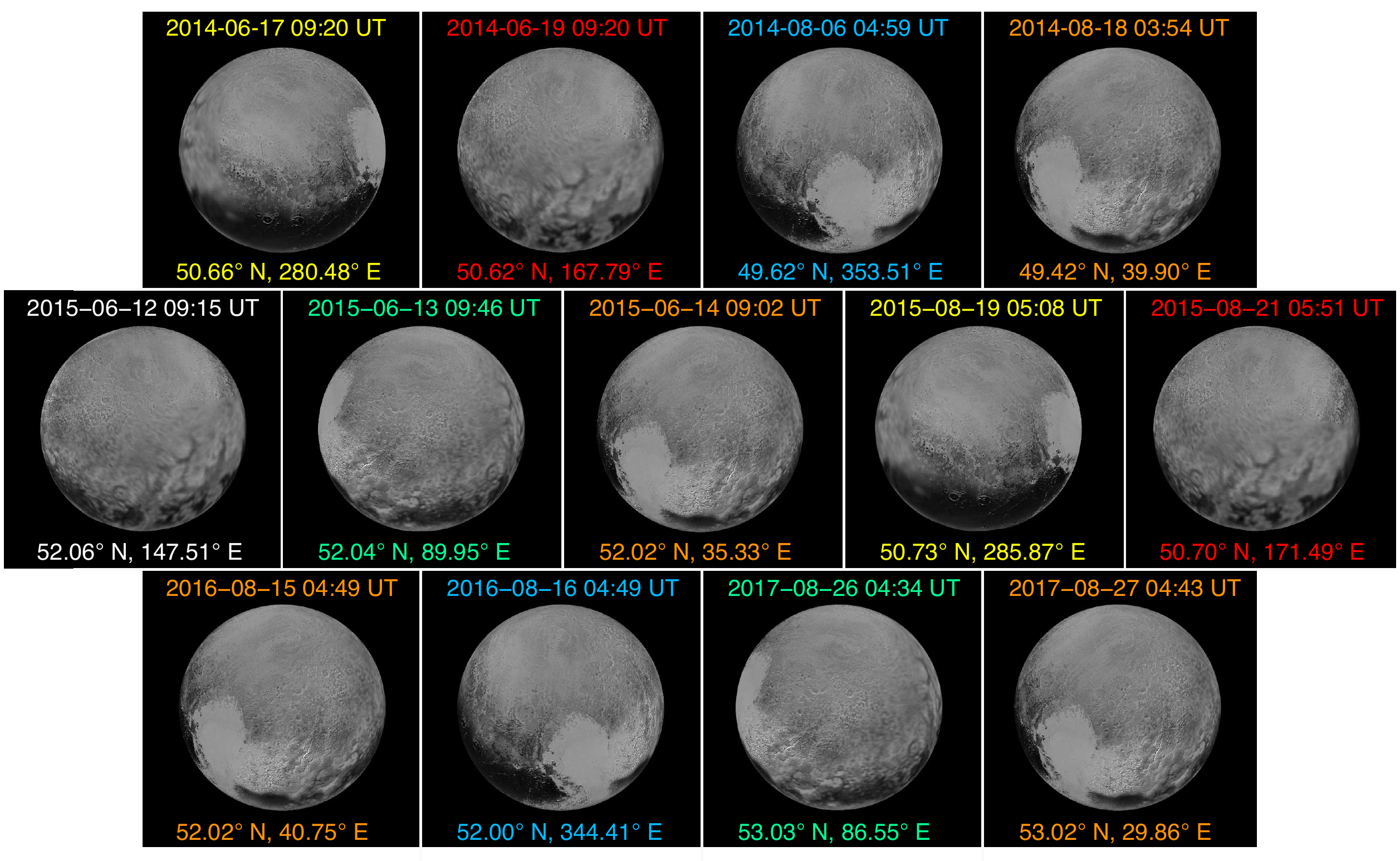}
\caption{Sub-observer hemispheres of Pluto as seen at the UT mid-time for each night. Sub-observer latitude and longitude, in the right hand rule coordinate system, are reported below each Pluto image. Text color indicates matched pairs and correspond to the same colors for the spectra in Figure~\ref{avgspec}. Four nights (indicated with orange text) are all within 4$^{\circ}$ of sub-observer latitude and 11$^{\circ}$ of sub-observer longitude, and are thus considered a ``matched quartet." There is no matching night for 2015-06-12. The image of Pluto was obtained by New Horizons and retrieved from the JPL Photojournal (\url{https://photojournal.jpl.nasa.gov/}).\label{subobshemis}}
\end{center}
\end{figure}

\doublespacing{We obtained near-infrared spectra of the blended Pluto/Charon binary with the TripleSpec cross-dispersed spectrograph (Wilson et al., 2004) on the Astrophysical Research Consortium's 3.5-meter telescope at the Apache Point Observatory (APO). With seeing typically on the order of 1", we used the 1.1" slit for all observations to reduce slit losses. Use of this slit resulted in an average resolving power of $\sim$3500 across the 0.91-2.47 $\mu$m wavelength range. Also, due to the typical seeing, the Pluto and Charon PSFs were always at least partially blended (the maximum elongation of Charon was $\sim$0.8" over the time period of these observations). At the expected seeing, Charon would always contribute to the spectra, so we were careful to ensure that the contribution was as consistent as possible by orienting the slit along the imaginary line connecting the centers of Pluto and Charon.

\begin{table}[h!]
\begin{center}
\textbf{Table~\ref{obscircum}}\\
Observation circumstances\\
\footnotesize
\begin{tabular}{cccccc}
\hline
UT date & $Ks$ band & Sub-observer & Sub-observer & Phase angle & Integration\\
mid-time & seeing (") & latitude ($^{\circ}$)$^a$ & longitude ($^{\circ}$)$^a$ & ($^{\circ}$) & time (min)\\
\hline
2014-06-17 09:20 & 0.66 & 50.66 & 280.48 & 0.52 & 130\\
2014-06-19 09:20 & 1.2 & 50.62 & 167.79 & 0.46 & 110\\
2014-08-06 04:59 & 1.0 & 49.62 & 353.51 & 0.94 & 180\\
2014-08-18 03:54 & 1.0 & 49.42 & 39.90 & 1.22 & 80\\
2015-06-12 09:15 & 1.6 & 52.06 & 147.51 & 0.72 & 100\\
2015-06-13 09:46 & 1.0 & 52.04 & 89.95 & 0.69 & 100\\
2015-06-14 09:02 & 1.1 & 52.02 & 35.33 & 0.67 & 180\\
2015-08-19 05:08 & 1.24 & 50.73 & 285.87 & 1.19 & 160\\
2015-08-21 05:51 & 2.0 & 50.70 & 171.49 & 1.23 & 80\\
2016-08-15 04:49 & 2.5 & 52.02 & 40.75 & 1.06 & 200\\
2016-08-16 04:49 & 1.27 & 52.00 & 344.41 & 1.08 & 120\\
2017-08-26 04:34 & 1.13 & 53.03 & 86.55 & 1.24 & 70\\
2017-08-27 04:43 & 1.13 & 53.02 & 29.86 & 1.26 & 180\\
\hline
\end{tabular}
\end{center}
\captionlistentry{}
\footnotesize{$^a$Right hand rule coordinate system.}
\label{obscircum}
\end{table}

\indent Pluto/Charon spectra were obtained using an ABBA slit dither pattern, with individual spectra of 300 seconds at both the A and B positions; this exposure time was long enough to see the trace of the unresolved binary system in each raw image, but short enough to avoid changes in the sky background. Due to the latitude of APO ($\sim$33$^{\circ}$N), spectra were made each night at airmasses between 1.7 and 2.0. Observations of a nearby G2V standard star, HD 176453, were made at least three times during each night (beginning, middle, and end), with one ABBA dither pattern per visit and 30 seconds per dither position. The same G2V standard was used on each night throughout the program. Dome flats with a bright quartz lamp were obtained once each night during twilight. No neon or argon lamp spectra were obtained.\\
\indent We observed Pluto/Charon on 13 nights between June 2014 and August 2017; observation circumstances for each night are available in Table~\ref{obscircum}. Specific nights were requested in June from 2014-2016 and August from 2015-2017 in order to obtain matched pairs of observations at similar sub-observer latitudes and longitudes, thus observing similar hemispheres of Pluto in each half of the pair. The observed Pluto hemisphere from each night is shown in Figure~\ref{subobshemis}. When requesting time for the observations, we aimed for matched pairs within $\pm$5$^{\circ}$ of sub-observer latitude and $\pm$10$^{\circ}$ of sub-observer longitude. These constraints were qualitative and were chosen in order to reduce viewing geometry effects while allowing for multiple observing windows each semester. In some cases, bad weather frustrated our efforts to create matched pairs over a $\sim$14-month interval. On one occasion we were unable to create a matched pair at all: the closest night to 2015-06-12 in sub-observer longitude is 2014-06-19, with a separation of $\sim$20$^{\circ}$, well outside our initial constraint. Conversely, by stretching the matched pair definition very slightly (expanding the sub-observer longitude constraint by $<$1$^{\circ}$), we were able to create a ``matched quartet," with each member of the quartet paired with every other member to create 6 unique pairs. This matched quartet spans the full observing window from 2014 to 2017, enabling a search for temporal changes on that particular hemisphere of Pluto over the full 3-year time span.}

\section{Analysis \& Results}
\begin{figure}[ht!]
\begin{center}
\includegraphics[scale=0.76,trim=0cm 0cm 0cm 0cm,clip=true]{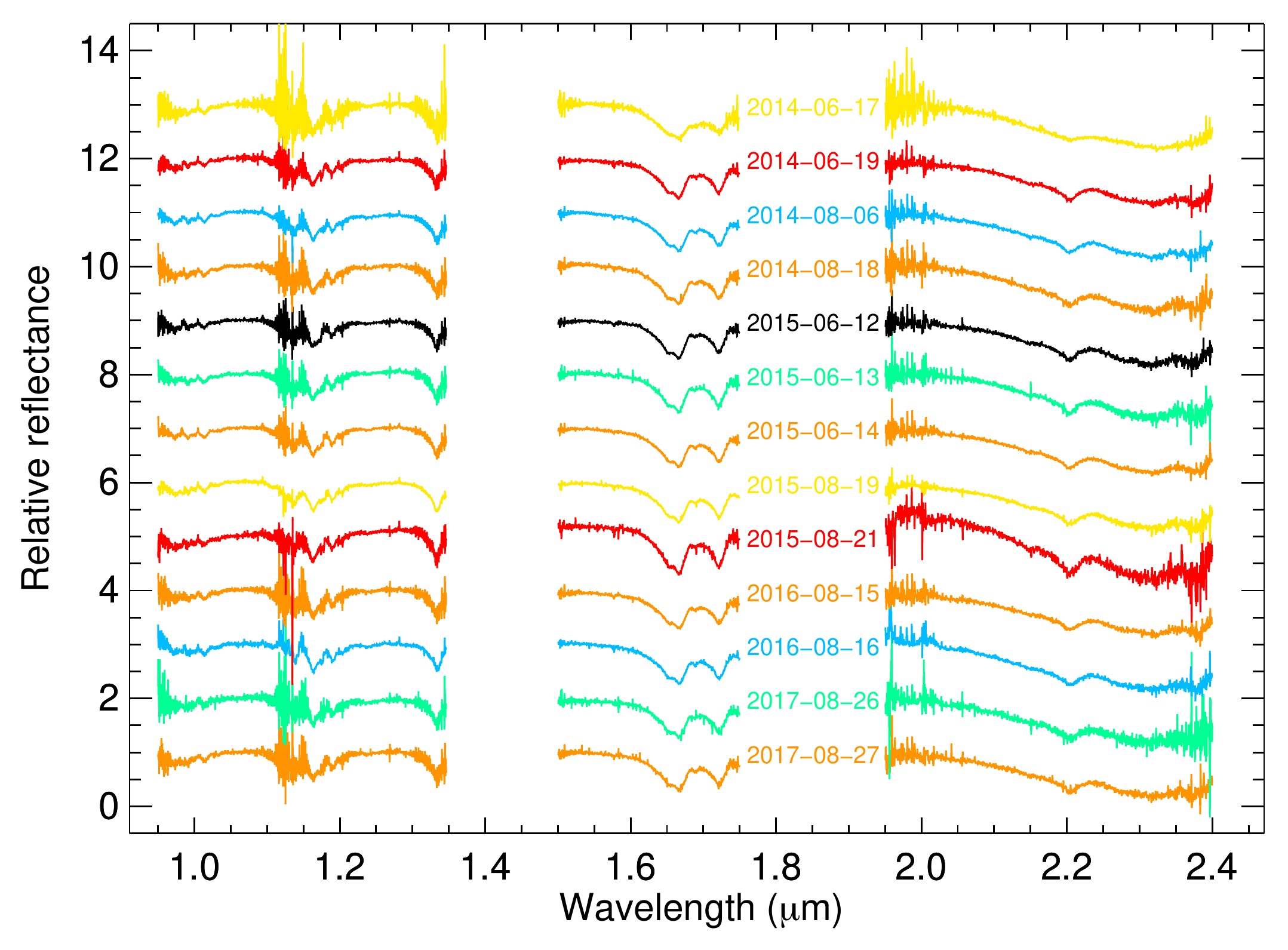}
\caption{Nightly averaged Pluto spectra color-coded by matched pair; see Figure~\ref{subobshemis} for the sub-observer hemispheres of Pluto corresponding to these colors. The spectra were normalized in the 1.03-1.08 $\mu$m region and are offset in the $y$-direction for clarity. In some cases, spectral changes are readily apparent, such as the depth of the 2.2 $\mu$m CH$_4$ feature for the yellow matched pair (2014-06-17 and 2015-08-19). Regions of strong telluric absorption between 1.35-1.5 and 1.75-1.95 $\mu$m, as well as the low-SNR edges of the spectra between 0.91-0.95 and 2.4-2.47 $\mu$m, were masked out. The UT date is presented for each spectrum in the masked-out region between 1.75 and 1.95 $\mu$m. The 2015-06-12 observations do not match up with any other night.\label{avgspec}}
\end{center}
\end{figure}

\doublespacing{Individual Pluto/Charon and HD 176453 spectra were reduced using the IDL-based {\it tspectool} package, which is based on the {\it spextool} package developed by Cushing et al. (2004) for the SpeX instrument at NASA's Infrared Telescope Facility (IRTF). The first step was to use {\it tspectool} to create a master flat field for identification of the spectral orders. Spectra were then run through the program, which performed wavelength calibration using airglow lines, AB pair subtraction, extraction aperture definition, trace fitting in each order, and spectral extraction within the defined aperture. User-input was occasionally required: Aperture and background widths were changed from the default value when necessary. Each output spectrum contained per-pixel information on wavelength, flux, and uncertainty. We used the {\it xcombspec} sub-package within {\it tspectool} to combine each order of the extracted spectra into a nightly average using a 3-$\sigma$ robust mean. The orders were then stitched together into a combined spectrum using the {\it xmergeorders} sub-package within {\it tspectool}. All nightly average spectra were sampled onto the same wavelength grid. The solar correction was performed by dividing the nightly average Pluto/Charon spectrum by the nightly average HD 176453 spectrum (see final corrected spectra in Figure~\ref{avgspec}). The solar-corrected spectra were normalized by the mean of the flat continuum region between 1.03 and 1.08 $\mu$m.\\
\indent We first attempted to divide the individual Pluto/Charon spectra in a given night by solar analogue spectra at matching airmasses, but upon examination of the telluric regions between 1.35-1.5 $\mu$m and 1.75-1.95 $\mu$m, we found that the choice of solar spectrum was irrelevant due to the strength of the telluric absorption. Pluto was observed at relatively high airmass each night, due to the latitude of APO, and water vapor was high in the summer months. This led to saturation in the 1.35-1.5 $\mu$m and 1.75-1.95 $\mu$m regions, and these ranges are masked out in Figure~\ref{avgspec}. Thus, we did not apply any additional telluric correction beyond division of the average Pluto/Charon spectra by the average solar analogue spectra.}

\begin{table}[ht!]
\begin{center}
\textbf{Table~\ref{bandregions}}\\
CH$_4$ absorption bands and continuum regions\\
\footnotesize
\begin{tabular}{cccccc}
\hline
Identifier & Short continuum ($\mu$m) & Absorption band ($\mu$m) & Long continuum ($\mu$m)\\
\hline
1.16 $\mu$m & 1.061 -- 1.080 & 1.161 -- 1.165 & 1.245 -- 1.265\\
1.19 $\mu$m & 1.061 -- 1.080 & 1.188 -- 1.191 & 1.245 -- 1.265\\
1.33 $\mu$m & 1.250 -- 1.280 & 1.330 -- 1.335 & 1.510 -- 1.540\\
1.66 $\mu$m & 1.510 -- 1.540 & 1.664 -- 1.668 & 1.980 -- 2.010\\
1.72 $\mu$m & 1.510 -- 1.540 & 1.718 -- 1.723 & 1.980 -- 2.010\\
\hline
\end{tabular}
\end{center}
\captionlistentry{}
\label{bandregions}
\end{table}

\begin{figure}[ht!]
\begin{center}
\includegraphics[scale=0.6,trim=0cm 0cm 0cm 0cm,clip=true]{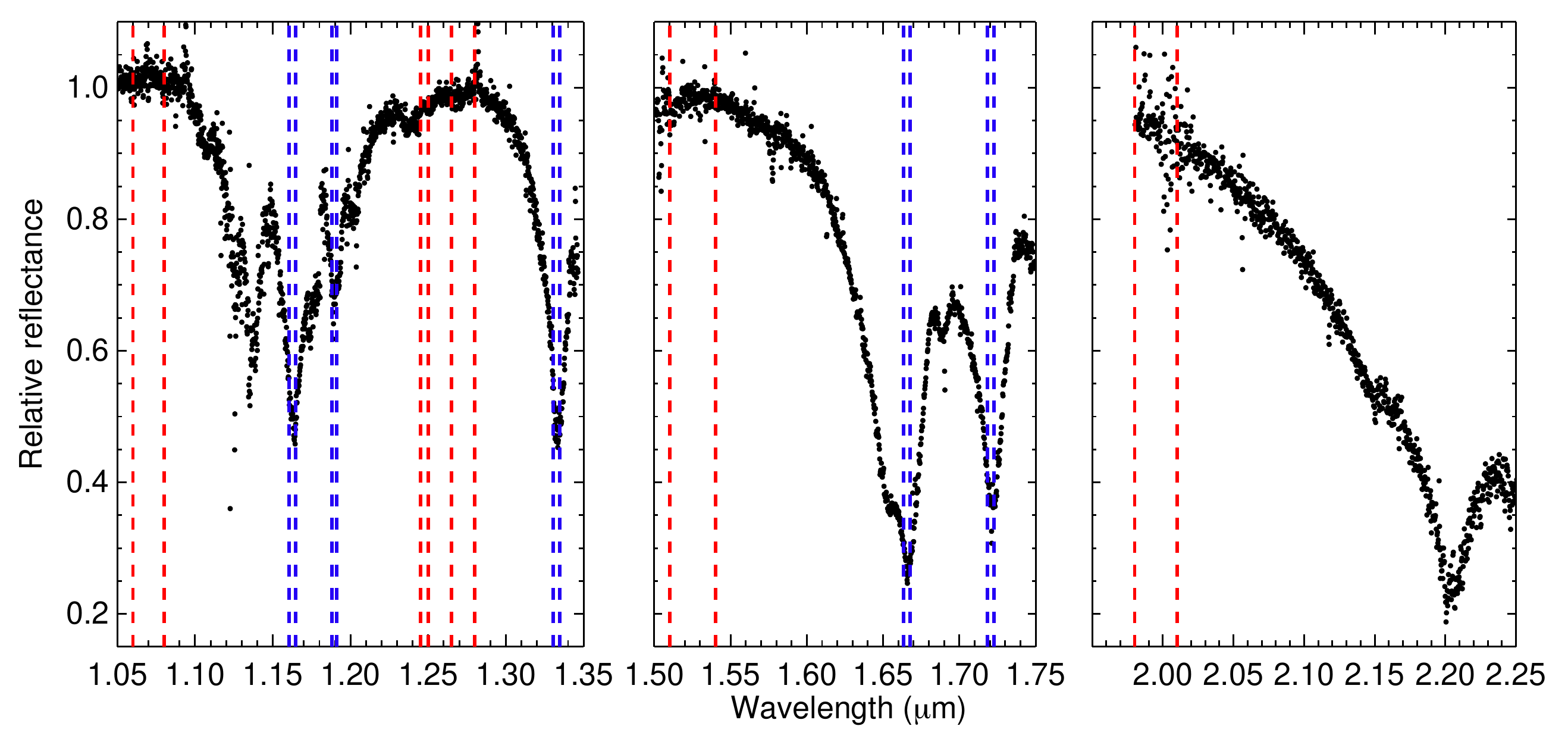}
\caption{Visualization of the CH$_4$ absorption band and continuum regions presented in Table~\ref{bandregions}, using the nightly average spectrum from 2015-08-19. Red dashed lines mark the boundaries of the continuum regions and the blue dashed lines mark the bands themselves. The regions of strong telluric absorption are omitted and the gaps between the separately plotted wavelength regions are not to-scale.\label{plotregions}}
\end{center}
\end{figure}

\begin{figure}[ht!]
\begin{center}
\includegraphics[scale=0.6,trim=0cm 0cm 0cm 0cm,clip=true]{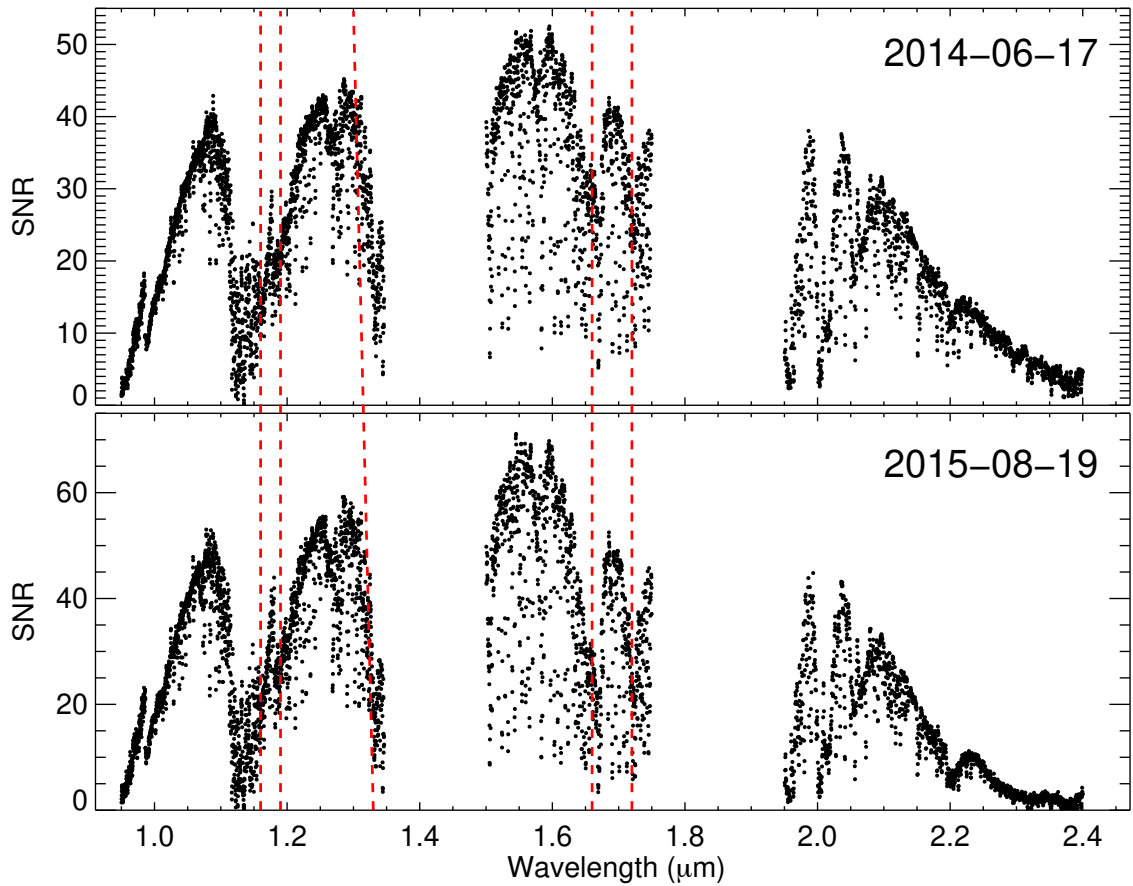}
\caption{Signal-to-noise ratio (SNR) of the nightly average spectra for 2014-06-17 and 2015-08-19. The regions of strong telluric absorption between 1.35-1.5 and 1.75-1.95 $\mu$m are omitted. Red dashed lines mark the positions of the CH$_4$ absorption bands considered in this work. The center of all the bands are at or above SNR=20 for both nights. See the text for a description of how the uncertainties were computed.\label{specSNR}}
\end{center}
\end{figure}

\doublespacing{To evaluate changes in surface composition over time, we computed integrated band areas for the 1.16, 1.19, 1.33, 1.66, and 1.72 $\mu$m CH$_4$ absorption bands in each of the corrected nightly spectra. The relative strengths of the 1.66 and 1.72 $\mu$m bands compared to the other bands and the higher signal-to-noise ratio (SNR) in this wavelength region (Figure~\ref{avgspec}) made them the most appropriate for this investigation. Band areas provide a useful metric for comparing ice properties between two or more spectra due to differences in abundance, concentration, and/or grain size (path length) and are straightforward to calculate and conceptualize. To calculate the band areas, three wavelength regions were identified for each absorption feature of interest (Table~\ref{bandregions}, Figure~\ref{plotregions}): the absorption feature itself, a continuum region at shorter wavelengths, and a continuum region at longer wavelengths. The continuum regions were not necessarily flat (zero slope), but were linear and did not contain any portion of a large absorption feature. All three regions were not necessarily contiguous. A line was fit to the points in both continuum regions using a least squares fitting routine, then the entire spectrum was normalized by the best-fit line. This sets the continua on either side of the band to 1, allowing for simple numerical integration of the area under the absorption feature (between the absorption feature and 0). Taking 1 minus this area resulted in the area of the absorption feature itself. The uncertainties on the linear fit to the continua were not propagated as part of the error analysis because they were negligible compared to the uncertainties on the points themselves.\\
\indent Following visual inspection, the uncertainties output by {\it tspectool} were found to be inconsistent with the magnitude of the scatter in the spectra. In order to get a better estimate of the uncertainties, we calculated two quantities in the continuum region between 1.03 and 1.08 $\mu$m in each normalized nightly spectrum: the standard deviation of the flux and the mean of the uncertainties. The standard deviation of the flux provided an estimate of the scatter about a region of the spectrum that could be modeled by a flat line. Dividing the first quantity by the latter, we computed the factor by which the uncertainties were underestimated by {\it tspectool}. This factor ranged from 71 to 372, depending on the spectrum; these large factors convincingly demonstrate how severely {\it tspectool} underestimated the uncertainties. We then multiplied all uncertainties by this factor when calculating the band areas and their uncertainties, which are presented in Table~\ref{bandareatab}. These more realistic uncertainties were also used to compute the SNRs presented in Figure~\ref{specSNR}.}

\begin{figure}[ht!]
\begin{center}
\includegraphics[scale=0.68,trim=0cm 0cm 0cm 0cm,clip=true]{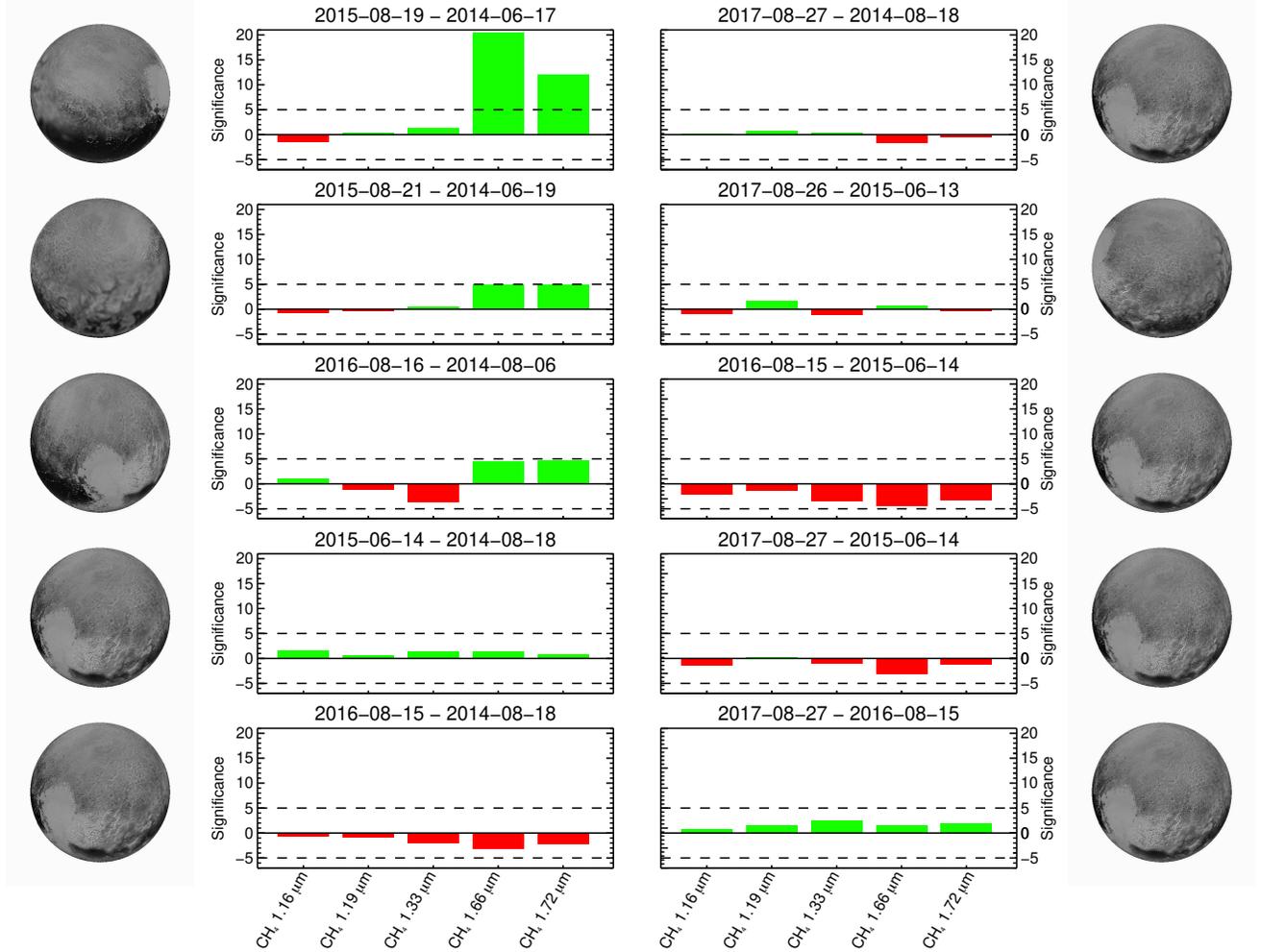}
\caption{Significance of band area differences for each matched pair (see Table~\ref{bandareatab}). The horizontal dashed lines indicate the $\pm$5-$\sigma$ significance thresholds. Calculations are the later date minus the earlier date; thus, green bars indicate that the band area increased over time and red bars indicate that the band area decreased over time. The approximate observed hemisphere of Pluto for each matched pair is presented to the side of each plot. The image of Pluto was obtained by New Horizons and retrieved from the JPL Photojournal (\url{https://photojournal.jpl.nasa.gov/}).\label{bandarea}}
\end{center}
\end{figure}

\begin{table}[ht!]
\begin{center}
\textbf{Table~\ref{bandareatab}}\\
CH$_4$ band areas\\
\footnotesize
\begin{tabular}{cccccc}
\hline
UT date & 1.16 $\mu$m & 1.19 $\mu$m & 1.33 $\mu$m & 1.66 $\mu$m & 1.72 $\mu$m\\
& band area (nm) & band area (nm) & band area (nm) & band area (nm) & band area (nm)\\
\hline
2015-08-19 & 1.932 $\pm$ 0.021 & 0.939 $\pm$ 0.020 & 1.995 $\pm$ 0.032 & 2.787 $\pm$ 0.013 & 2.377 $\pm$ 0.022\\
2014-06-17 & 1.985 $\pm$ 0.029 & 0.927 $\pm$ 0.026 & 1.935 $\pm$ 0.035 & 2.416 $\pm$ 0.013 & 1.997 $\pm$ 0.023\\
$\Delta$ & -0.053 $\pm$ 0.036 & 0.012 $\pm$ 0.033 & 0.060 $\pm$ 0.047 & 0.371 $\pm$ 0.018 & 0.380 $\pm$ 0.032\\
\hline
2015-08-21 & 1.822 $\pm$ 0.031 & 0.808 $\pm$ 0.031 & 1.920 $\pm$ 0.051 & 2.887 $\pm$ 0.019 & 2.511 $\pm$ 0.032\\
2014-06-19 & 1.853 $\pm$ 0.023 & 0.821 $\pm$ 0.022 & 1.893 $\pm$ 0.029 & 2.774 $\pm$ 0.013 & 2.324 $\pm$ 0.020\\
$\Delta$ & -0.031 $\pm$ 0.039 & -0.013 $\pm$ 0.038 & 0.027 $\pm$ 0.059 & 0.113 $\pm$ 0.023 & 0.187 $\pm$ 0.038\\
\hline
2016-08-16 & 1.881 $\pm$ 0.027 & 0.905 $\pm$ 0.024 & 1.729 $\pm$ 0.035 & 2.804 $\pm$ 0.015 & 2.447 $\pm$ 0.021\\
2014-08-06 & 1.841 $\pm$ 0.023 & 0.944 $\pm$ 0.023 & 1.909 $\pm$ 0.033 & 2.711 $\pm$ 0.014 & 2.299 $\pm$ 0.024\\
$\Delta$ & 0.040 $\pm$ 0.035 & -0.039 $\pm$ 0.033 & -0.180 $\pm$ 0.048 & 0.093 $\pm$ 0.021 & 0.148 $\pm$ 0.032\\
\hline
2015-06-14 & 1.863 $\pm$ 0.017 & 0.869 $\pm$ 0.017 & 1.889 $\pm$ 0.022 & 2.700 $\pm$ 0.011 & 2.303 $\pm$ 0.017\\
2014-08-18 & 1.821 $\pm$ 0.019 & 0.853 $\pm$ 0.020 & 1.840 $\pm$ 0.029 & 2.676 $\pm$ 0.012 & 2.280 $\pm$ 0.021\\
$\Delta$ & 0.042 $\pm$ 0.025 & 0.016 $\pm$ 0.026 & 0.049 $\pm$ 0.036 & 0.024 $\pm$ 0.016 & 0.023 $\pm$ 0.027\\
\hline
2016-08-15 & 1.801 $\pm$ 0.024 & 0.829 $\pm$ 0.022 & 1.756 $\pm$ 0.031 & 2.614 $\pm$ 0.016 & 2.213 $\pm$ 0.022\\
2014-08-18 & 1.821 $\pm$ 0.019 & 0.853 $\pm$ 0.020 & 1.840 $\pm$ 0.029 & 2.676 $\pm$ 0.012 & 2.280 $\pm$ 0.021\\
$\Delta$ & -0.020 $\pm$ 0.031 & -0.024 $\pm$ 0.030 & -0.084 $\pm$ 0.042 & -0.062 $\pm$ 0.020 & -0.067 $\pm$ 0.030\\
\hline
2017-08-27 & 1.824 $\pm$ 0.022 & 0.875 $\pm$ 0.020 & 1.851 $\pm$ 0.026 & 2.645 $\pm$ 0.013 & 2.268 $\pm$ 0.021\\
2014-08-18 & 1.821 $\pm$ 0.019 & 0.853 $\pm$ 0.020 & 1.840 $\pm$ 0.029 & 2.676 $\pm$ 0.012 & 2.280 $\pm$ 0.021\\
$\Delta$ & 0.003 $\pm$ 0.029 & 0.022 $\pm$ 0.028 & 0.011 $\pm$ 0.039 & -0.031 $\pm$ 0.018 & 0.012 $\pm$ 0.030\\
\hline
2017-08-26 & 1.746 $\pm$ 0.027 & 0.813 $\pm$ 0.026 & 1.713 $\pm$ 0.049 & 2.701 $\pm$ 0.014 & 2.325 $\pm$ 0.022\\
2015-06-13 & 1.776 $\pm$ 0.016 & 0.766 $\pm$ 0.016 & 1.789 $\pm$ 0.021 & 2.660 $\pm$ 0.010 & 2.275 $\pm$ 0.017\\
$\Delta$ & -0.030 $\pm$ 0.031 & 0.047 $\pm$ 0.031 & -0.076 $\pm$ 0.053 & 0.041 $\pm$ 0.017 & 0.050 $\pm$ 0.028\\
\hline
2016-08-15 & 1.801 $\pm$ 0.024 & 0.829 $\pm$ 0.022 & 1.756 $\pm$ 0.031 & 2.614 $\pm$ 0.016 & 2.213 $\pm$ 0.022\\
2015-06-14 & 1.863 $\pm$ 0.017 & 0.869 $\pm$ 0.017 & 1.889 $\pm$ 0.022 & 2.700 $\pm$ 0.011 & 2.303 $\pm$ 0.017\\
$\Delta$ & -0.062 $\pm$ 0.029 & -0.040 $\pm$ 0.028 & -0.133 $\pm$ 0.038 & -0.086 $\pm$ 0.019 & -0.090 $\pm$ 0.028\\
\hline
2017-08-27 & 1.824 $\pm$ 0.022 & 0.875 $\pm$ 0.020 & 1.851 $\pm$ 0.026 & 2.645 $\pm$ 0.013 & 2.268 $\pm$ 0.021\\
2015-06-14 & 1.863 $\pm$ 0.017 & 0.869 $\pm$ 0.017 & 1.889 $\pm$ 0.022 & 2.700 $\pm$ 0.011 & 2.303 $\pm$ 0.017\\
$\Delta$ & -0.039 $\pm$ 0.028 & 0.006 $\pm$ 0.026 & -0.038 $\pm$ 0.034 & -0.055 $\pm$ 0.017 & -0.035 $\pm$ 0.027\\
\hline
2017-08-27 & 1.824 $\pm$ 0.022 & 0.875 $\pm$ 0.020 & 1.851 $\pm$ 0.026 & 2.645 $\pm$ 0.013 & 2.268 $\pm$ 0.021\\
2016-08-15 & 1.801 $\pm$ 0.024 & 0.829 $\pm$ 0.022 & 1.756 $\pm$ 0.031 & 2.614 $\pm$ 0.016 & 2.213 $\pm$ 0.022\\
$\Delta$ & 0.023 $\pm$ 0.033 & 0.046 $\pm$ 0.030 & 0.095 $\pm$ 0.040 & 0.031 $\pm$ 0.021 & 0.055 $\pm$ 0.030\\
\hline
\end{tabular}
\end{center}
\captionlistentry{}
\label{bandareatab}
\end{table}

\doublespacing{After computing the band areas and uncertainties for the five CH$_4$ bands of interest in each spectrum, we calculated the band area differences for each matched pair. Differences were computed as the later date minus the earlier date, so positive values indicate an increase in the band area with time. The band area differences and propagated errors are presented on the lines marked with $\Delta$s in Table~\ref{bandareatab}, below the band areas for the components of each matched pair. Figure~\ref{bandarea} presents the significance of the band area differences and an image of Pluto's sub-observer hemisphere for each matched pair. Only those short-term changes detected at $\pm$5-$\sigma$ were considered statistically significant. The only significant change was detected between 2014-06-17 and 2015-08-19, centered on a sub-observer longitude of $\sim$280$^{\circ}$, in the right hand rule coordinate system used as the IAU standard, and described in Zangari (2015). No other matched pair showed a significant change over the corresponding time period.\\
\indent This work focused solely on the 5 CH$_4$ bands mentioned previously due to difficulties in evaluating some of the weak and very strong CH$_4$ absorption features, as well as the N$_2$ and CO (carbon monoxide) features. Specifically, we did not consider the 0.97, 0.99, 2.20, or 2.32 $\mu$m CH$_4$ absorption features in this analysis. The 0.97 and 0.99 $\mu$m features were too weak and too close to the low-SNR edges of the spectra (Figure~\ref{specSNR}), while the 2.20 and 2.32 $\mu$m features were too strong, preventing appropriate continuum regions from being identified (Figure~\ref{avgspec}). Strong CH$_4$ bands on either side of the 2.35 $\mu$m CO band prevented evaluation of that feature; any variations noticed in this CO band would have been challenging to disentangle from variations in CH$_4$. The CO band at 1.58 $\mu$m was too weak to be useful for this investigation. N$_2$ ice has one feature in this wavelength range at 2.15 $\mu$m, but the feature was relatively weak and the SNR of the spectra at longer wavelengths frustrated efforts to quantify any changes in band area over time.}

\section{Discussion}
\doublespacing{Due to scheduling and weather, the majority of matched pairs were obtained of the anti-Charon hemisphere, home to the bright, volatile-rich Sputnik Planitia. However, the only sub-observer hemisphere with a statistically significant band area increase, centered on $\sim$280$^{\circ}$, included only a small portion of Sputnik Planitia but a large portion of the low-albedo, volatile-poor Cthulhu Macula (e.g., Grundy et al., 2016; Protopapa et al., 2017). The sub-observer hemisphere for the 2015-08-19/2014-06-17 matched pair was unique among all matched pairs and contained the largest fraction of Cthulhu Macula. We explore possible explanations for the observed changes in the remainder of this section.\\
\indent We can rule out the influence of phase angle effects and differences in spectral slope on our results. In general, viewing a slab of ice at an oblique angle results in a longer path length and thus a stronger absorption band (e.g., Pitman et al., 2017), so reducing the effects of viewing geometry is critical for this kind of comparative investigation. The timing of observations in June and August, roughly one month before and after opposition, respectively, resulted in comparable phase angles between the components of each matched pair (Table~\ref{obscircum}), reducing variations in band strength due to phase angle effects. Additionally, the phase angles of all observations fall on the linear portion of the phase curve, avoiding the non-linear opposition surge within phase angles of 0.1$^{\circ}$ (e.g., Verbiscer et al., 2020; Buratti et al., 2021; Olkin et al., 2021).\\
\indent We can also rule out limb darkening effects on the results. Previous laboratory work shows that small changes in emission angle can result in changes in band depth (e.g., Gradie et al., 1980). The sub-observer hemisphere centered on $\sim$280$^{\circ}$ presents a small portion of the volatile-rich Sputnik Planitia on the limb, with a slightly larger portion in the later component. While the matched pair method significantly lessens the effect of limb darkening on the results, it is possible that small changes in sub-observer longitude could still result in detectable changes. In lieu of CH$_4$ band area as a function of emission angle in Sputnik Planitia, we compare the band area differences of a few matched pairs to evaluate the possibility that limb darkening is responsible for the reported band area increase in the 2015-08-19/2014-06-17 pair. For this pair, the difference in sub-observer longitude was $\sim$5$^{\circ}$. Two matched pairs have larger differences in sub-observer longitude: 2016-08-16/2014-08-06 at $\sim$9$^{\circ}$ and 2017-08-27/2014-08-18 at $\sim$10$^{\circ}$ (third panel on the left and the top right panel in Figure~\ref{bandarea}, respectively). The sub-observer hemispheres for these matched pairs present roughly comparable areas of Sputnik Planitia on their limbs compared to the 2015-08-19/2014-06-17 pair, yet the band area differences do not rise to the 5-$\sigma$ level. All but one matched pair have portions of Sputnik Planitia on the limb of their sub-observer hemispheres, and none of these pairs show a significant change in band area either. Given this, we were able to rule out limb darkening as the cause of the CH$_4$ band area increase between 2014-06-17 and 2015-08-19.\\
\indent We also evaluated the effects of small differences in spectral slope between matched pair components caused by slit orientations not aligned with the parallactic angle. Atmospheric dispersion of light from Pluto and Charon in the vertical direction along the parallactic angle can result in slit losses at some, but not all, wavelengths. Thus, aligning the slit such that it is along the parallactic angle reduces wavelength-dependent slit losses. In this investigation, however, the slit was always aligned along the Pluto-Charon line, as described in the Observations section, which rarely, if ever, aligned with the parallactic angle. Pluto and Charon are doubly synchronous, resulting in a comparable angle between the two for each component of the matched pair. However, different hour angles in June and August led to different slit orientations with respect to the parallactic angle. To evaluate the effect of potential spectral slope differences on the band area calculations, we took a CH$_4$ band and its continuum regions from one of the averaged spectra, added linear components with varying degrees of positive and negative slopes, and calculated the band area for each value of the slope. The standard deviation of the band areas was only 0.7\% of the average. Therefore, we conclude that our method of calculating band areas is robust against the effects of varying spectral slope.\\
\indent The increase in CH$_4$ band areas in the spectra of one unique sub-observer hemisphere between June 2014 and August 2015 points to real short-term changes in Pluto's surface composition over this time frame. This increase in CH$_4$ band areas could be due to (1) an increase in CH$_4$ grain size over time, (2) deposition of CH$_4$ from the atmosphere as Pluto moves away from the Sun, or (3) preferential sublimation of N$_2$ ice from the north polar region as the sub-solar latitude migrates northward. We evaluate each potential explanation below.
\begin{enumerate}[listparindent=\parindent]
\item {\it Increase in CH$_4$ grain size}: Sintering, when adjacent ice grains combine under the effects of heat or pressure without liquefaction, is a possible mechanism for the growth of CH$_4$ grains on the surface of Pluto (e.g., Molaro et al., 2018). Modeling work for sintering of N$_2$ and CH$_4$ ices on the surface of Triton indicates that grains with diameters on the order of 1-10 $\mu$m could form through sintering in 1 year at a temperature of 45 K, depending on the porosity of the ice layer (Eluszkiewicz, 1991). However, Hapke modeling (Hapke, 2012) of spectra obtained across Pluto's surface by New Horizons suggests that the absorption features primarily result from CH$_4$-rich grains 100s or even 1000s of microns in diameter (Protopapa et al., 2017). Sintering of grains of this size is negligible over the course of a single year, and it is unlikely that the creation of grains an order of magnitude smaller would significantly increase the strength of the CH$_4$ absorption features. For these reasons, we do not favor this explanation.

\item {\it Deposition of CH$_4$}: We do not observe a change in all CH$_4$ absorption bands because some bands are inherently stronger than others. The features observed in Pluto's near-infrared spectrum are due to vibrational transitions caused by absorption of photons of specific energies; some of these transitions are more likely to occur than others, corresponding to higher absorption coefficients. Inherently stronger bands, those with higher absorption coefficients, are also more sensitive to changes in the abundance of the ice. The strongest CH$_4$ bands considered in this investigation, the 1.66 and 1.72 $\mu$m features, were the only two that showed a statistically significant increase in band area on the single sub-observer hemisphere where changes were detected.\\
\indent Of the three volatile ices on Pluto's surface (N$_2$, CO, and CH$_4$), CH$_4$ is the least volatile (e.g., Fray and Schmitt, 2009). However, assuming the primary driver of a change in atmospheric pressure is a change in atmospheric temperature, deposition of N$_2$ onto Pluto's surface would exceed that of CH$_4$. According to Fray and Schmitt (2009), who use the same functional form to describe the pressure as a function of temperature for N$_2$ and CH$_4$, a change from 50 K to 40 K would result in a decrease in N$_2$ pressure nearly 3 orders of magnitude larger than that of CH$_4$. Coupled with the fact that Pluto's atmosphere is composed predominantly of N$_2$ (the CH$_4$:N$_2$ mixing ratio was determined by New Horizons to be 0.008; Gladstone et al., 2016), any detectable change in CH$_4$ absorption due to deposition would be accompanied by detectable changes in the CH$_4$ band shifts due to deposition of N$_2$.\\
\indent Specifically, such short-term atmospheric evolution would result in a detectable blueshift of the CH$_4$ band centers (e.g., Quirico and Schmitt, 1997; Protopapa et al., 2015). The 2.15 $\mu$m N$_2$ ice feature was very weak in all of our Pluto spectra (Figure~\ref{avgspec}), which made it impossible to calculate the N$_2$ band area and quantify temporal changes in N$_2$ ice directly. The N$_2$ ice was instead investigated indirectly through the shift in the stronger CH$_4$ absorption features. Dilution of pure CH$_4$ ice with N$_2$, or pure N$_2$ ice with CH$_4$, alters the dipole moment of CH$_4$ and shifts the central wavelength of the CH$_4$ absorption features (e.g., Protopapa et al., 2015). Specifically, a larger blueshift is indicative of a higher N$_2$ concentration, with the caveat that only small concentrations of each ice are miscible in the other (e.g., Prokhvatilov and Yantsevich, 1983).\\
\indent For these observations, we were less interested in determining the exact mixing ratio between CH$_4$ and N$_2$, and more interested in the relative band shifts over the time period of each matched pair. Our first attempt at calculating the relative shifts involved comparison of the nightly average Pluto spectra to a Hapke model (Hapke, 2012) of pure CH$_4$ ice generated using the optical constants of Grundy et al. (2002). Positive and negative shifts were applied to the model with respect to the nightly average spectrum, with the $\chi^2$ computed at each shift. Both the model and the spectra were sampled onto the same wavelength grid as the nightly average spectra and the shifts were calculated by multiplying the integer number of shifted wavelength bins by the average bin size in the wavelength range considered. The shift was calculated separately for each CH$_4$ band considered in this investigation because the behavior of each band is different for a given N$_2$ concentration (Protopapa et al., 2015). The relative shift of each CH$_4$ band was computed as the later date minus the earlier date, meaning negative values correspond to a relative blueshift over time. Uncertainties (1-$\sigma$) were determined by fitting a parabola to the $\chi^2$ values calculated as a function of the applied shift (in microns), determining the minimum of the parabola, and calculating the shifts corresponding to ($\chi^2_{min}$+1), as described in Tegler et al. (2008).\\
\indent As a check on this procedure, we performed a second analysis that compared the CH$_4$ bands of each matched pair directly. Consider that the most accurate results from the correlation method would involve a model perfectly to the data, such that the minimum $\chi^2$ would be achieved by simply aligning the band minima. We therefore decided to eliminate the comparison to a model and instead compared the band minima from each component of the matched pair directly. Fourier fits were performed over the wavelength regions for each band listed in Table~\ref{bandfits}, using the IDL routine {\it fourfit}\renewcommand{\thefootnote}{\arabic{footnote}}\footnote{The {\it fourfit} routine is part of Marc Buie's IDL Library (\url{https://www.boulder.swri.edu/~buie/idl/pro/fourfit.html}).}. Each Fourier fit included 50 terms and the wavelength regions were large enough to include more data points than variables. The positions of the minima were then calculated from the Fourier fits and a 100-trial Monte Carlo was performed to determine the uncertainty on each minimum. Gaussian random noise was added to the spectrum based on the uncertainties reported from {\it tspectool} multiplied by the ratio of the standard deviation of the data to the mean of the uncertainties between 1.03 and 1.08 $\mu$m. The band shift uncertainties are the standard deviations of the minima from the 100 Monte Carlo trials. The minimum of the earlier night of each matched pair was then subtracted from the minimum of the later night, with uncertainties determined through error propagation.

\begin{table}[ht!]
\begin{center}
\textbf{Table~\ref{bandfits}}\\
CH$_4$ absorption band Fourier fit regions\\
\footnotesize
\begin{tabular}{cc}
\hline
Identifier & Fit region ($\mu$m)\\
\hline
1.16 $\mu$m & 1.154 -- 1.173\\
1.19 $\mu$m & 1.180 -- 1.200\\
1.33 $\mu$m & 1.300 -- 1.345\\
1.66 $\mu$m & 1.603 -- 1.700\\
1.72 $\mu$m & 1.695 -- 1.740\\
\hline
\end{tabular}
\end{center}
\captionlistentry{}
\label{bandfits}
\end{table}

\indent Comparison of the results of each method revealed glaring discrepancies. For instance, the 1.66 and 1.72 $\mu$m bands were found to be blueshifted with a significance $>$5-$\sigma$ between 2014-06-17 and 2015-08-19 using the correlation method. However, the minimum comparison method reported no shift between the two dates for the 1.72 $\mu$m band and a non-significant shift ($<$1-$\sigma$) for the 1.66 $\mu$m band. Examination of the 1.72 $\mu$m bands by-eye (Figure~\ref{bandshift}) supports the minimum comparison method. The noise in the spectra appears to affect the band shift calculations for the correlation method, so we instead trust the minimum comparison method and present these results in Table~\ref{bandshifttab}. The minimum comparison method results in no relative shifts at the $>$5-$\sigma$ level, and only two above the 3-$\sigma$ level for the 1.33 $\mu$m band. The matched pairs in question are 2015-08-19/2014-06-17 (anti-Sputnik hemisphere) and 2016-08-16/2014-08-06 (Sputnik hemisphere). The 1.33 $\mu$m band is directly adjacent to a region of strong telluric absorption (1.35-1.5 $\mu$m) and examination of the bands within each matched pair revealed that one member of each pair had low signal-to-noise, resulting in a questionable minimum determination. The lack of detectable CH$_4$ band center shifts means it is unlikely that CH$_4$ and N$_2$ deposited onto the surface of Pluto between 2014 and 2015.\\
\indent Additional evidence against the deposition of CH$_4$ and N$_2$ onto the surface is that there was no observed decrease in Pluto's atmospheric pressure between 2014 and 2015. In fact, stellar occultation results show that the atmospheric pressure continued to increase until at least 2016 (Meza et al., 2019). These measurements suggest that the primary constituent of Pluto's atmosphere, N$_2$, did not deposit onto the surface in sufficient quantity for spectroscopic detection between 2014 and 2015, indicating that significant deposition of CH$_4$ also did not occur during this time frame.

\begin{table}[ht!]
\begin{center}
\textbf{Table~\ref{bandshifttab}}\\
CH$_4$ band center shifts\\
\footnotesize
\begin{tabular}{cccccc}
\hline
UT date & 1.16 $\mu$m & 1.19 $\mu$m & 1.33 $\mu$m & 1.66 $\mu$m & 1.72 $\mu$m\\
& band min. ($\mu$m) & band min. ($\mu$m) & band min. ($\mu$m) & band min. ($\mu$m) & band min. ($\mu$m)\\
\hline
2015-08-19 & 1.16351 $\pm$ 0.00016 & 1.17995 $\pm$ 0.0040 & 1.33406 $\pm$ 0.00013 & 1.66700 $\pm$ 0.00014 & 1.72192 $\pm$ 0.00013\\
2014-06-17 & 1.16298 $\pm$ 0.00025 & 1.18992 $\pm$ 0.00028 & 1.33492 $\pm$ 0.00016 & 1.66721 $\pm$ 0.00023 & 1.72192 $\pm$ 0.00023\\
$\Delta$ & 5.3 $\pm$ 3.0 \AA & -99.7 $\pm$ 40.2 \AA & -8.6 $\pm$ 2.1 \AA & -2.1 $\pm$ 2.7 \AA & 0.0 $\pm$ 2.6 \AA\\
\hline
2015-08-21 & 1.16298 $\pm$ 0.00026 & 1.18957 $\pm$ 0.00030 & 1.33389 $\pm$ 0.00015 & 1.66700 $\pm$ 0.00020 & 1.72149 $\pm$ 0.00016\\
2014-06-19 & 1.16263 $\pm$ 0.00020 & 1.18922 $\pm$ 0.00028 & 1.33371 $\pm$ 0.00011 & 1.66657 $\pm$ 0.00014 & 1.72127 $\pm$ 0.00012\\
$\Delta$ & 3.5 $\pm$ 3.3 \AA & 3.5 $\pm$ 4.1 \AA & 1.8 $\pm$ 1.9 \AA & 4.3 $\pm$ 2.4 \AA & 2.2 $\pm$ 2.0 \AA\\
\hline
2016-08-16 & 1.16386 $\pm$ 0.00017 & 1.19027 $\pm$ 0.00023 & 1.33458 $\pm$ 0.00016 & 1.66721 $\pm$ 0.00015 & 1.72192 $\pm$ 0.00015\\ 
2014-08-06 & 1.16351 $\pm$ 0.00017 & 1.18992 $\pm$ 0.00020 & 1.33389 $\pm$ 0.00016 & 1.66721 $\pm$ 0.00016 & 1.72170 $\pm$ 0.00014\\
$\Delta$ & 3.5 $\pm$ 2.4 \AA & 3.5 $\pm$ 3.0 \AA & 6.9 $\pm$ 2.3 \AA & 0.0 $\pm$ 2.2 \AA & 2.2 $\pm$ 2.1 \AA\\
\hline
2015-06-14 & 1.16333 $\pm$ 0.00011 & 1.18992 $\pm$ 0.00013 & 1.33354 $\pm$ 0.00010 & 1.66700 $\pm$ 0.00011 & 1.72149 $\pm$ 0.00011\\
2014-08-18 & 1.16351 $\pm$ 0.00013 & 1.18992 $\pm$ 0.00017 & 1.33371 $\pm$ 0.00012 & 1.66700 $\pm$ 0.00015 & 1.72170 $\pm$ 0.00015\\
$\Delta$ & -1.8 $\pm$ 1.7 \AA & 0.0 $\pm$ 2.1 \AA & -1.7 $\pm$ 1.6 \AA & 0.0 $\pm$ 1.9 \AA & -2.1 $\pm$ 1.9 \AA\\
\hline
2016-08-15 & 1.16368 $\pm$ 0.00017 & 1.18992 $\pm$ 0.00018 & 1.33371 $\pm$ 0.00013 & 1.66721 $\pm$ 0.00014 & 1.72170 $\pm$ 0.00016\\
2014-08-18 & 1.16351 $\pm$ 0.00013 & 1.18992 $\pm$ 0.00017 & 1.33371 $\pm$ 0.00012 & 1.66700 $\pm$ 0.00015 & 1.72170 $\pm$ 0.00015\\
$\Delta$ & 1.7 $\pm$ 2.1 \AA & 0.0 $\pm$ 2.5 \AA & 0.0 $\pm$ 1.8 \AA & 2.1 $\pm$ 2.1 \AA & 0.0 $\pm$ 2.2 \AA\\
\hline
2017-08-27 & 1.16351 $\pm$ 0.00014 & 1.19009 $\pm$ 0.00015 & 1.33389 $\pm$ 0.00012 & 1.66721 $\pm$ 0.00014 & 1.72192 $\pm$ 0.00013\\ 
2014-08-18 & 1.16351 $\pm$ 0.00013 & 1.18992 $\pm$ 0.00017 & 1.33371 $\pm$ 0.00012 & 1.66700 $\pm$ 0.00015 & 1.72170 $\pm$ 0.00015\\ 
$\Delta$ & 0.0 $\pm$ 1.9 \AA & 1.7 $\pm$ 2.3 \AA & 1.8 $\pm$ 1.7 \AA & 2.1 $\pm$ 2.1 \AA & 2.2 $\pm$ 2.0 \AA\\
\hline
2017-08-26 & 1.16351 $\pm$ 0.00015 & 1.19009 $\pm$ 0.00021 & 1.33389 $\pm$ 0.00018 & 1.66700 $\pm$ 0.00013 & 1.72149 $\pm$ 0.00013\\
2015-06-13 & 1.16316 $\pm$ 0.00012 & 1.18957 $\pm$ 0.00015 & 1.33337 $\pm$ 0.00009 & 1.66678 $\pm$ 0.00012 & 1.72127 $\pm$ 0.00010\\
$\Delta$ & 3.5 $\pm$ 1.9 \AA & 5.2 $\pm$ 2.6 \AA & 5.2 $\pm$ 2.0 \AA & 2.2 $\pm$ 1.8 \AA & 2.2 $\pm$ 1.6 \AA\\
\hline
2016-08-15 & 1.16368 $\pm$ 0.00017 & 1.18992 $\pm$ 0.00018 & 1.33371 $\pm$ 0.00013 & 1.66721 $\pm$ 0.00014 & 1.72170 $\pm$ 0.00016\\
2015-06-14 & 1.16333 $\pm$ 0.00011 & 1.18992 $\pm$ 0.00013 & 1.33354 $\pm$ 0.00010 & 1.66700 $\pm$ 0.00011 & 1.72149 $\pm$ 0.00011\\
$\Delta$ & 3.5 $\pm$ 2.0 \AA & 0.0 $\pm$ 2.2 \AA & 1.7 $\pm$ 1.6 \AA & 2.1 $\pm$ 1.8 \AA & 2.1 $\pm$ 1.9 \AA\\
\hline
2017-08-27 & 1.16351 $\pm$ 0.00014 & 1.19009 $\pm$ 0.00015 & 1.33389 $\pm$ 0.00012 & 1.66721 $\pm$ 0.00014 & 1.72192 $\pm$ 0.00013\\
2015-06-14 & 1.16333 $\pm$ 0.00011 & 1.18992 $\pm$ 0.00013 & 1.33354 $\pm$ 0.00010 & 1.66700 $\pm$ 0.00011 & 1.72149 $\pm$ 0.00011\\
$\Delta$ & 1.8 $\pm$ 1.8 \AA & 1.7 $\pm$ 2.0 \AA & 3.5 $\pm$ 1.6 \AA & 2.1 $\pm$ 1.8 \AA & 4.3 $\pm$ 1.7 \AA\\
\hline
2017-08-27 & 1.16351 $\pm$ 0.00014 & 1.19009 $\pm$ 0.00015 & 1.33389 $\pm$ 0.00012 & 1.66721 $\pm$ 0.00014 & 1.72192 $\pm$ 0.00013\\
2016-08-15 & 1.16368 $\pm$ 0.00017 & 1.18992 $\pm$ 0.00018 & 1.33371 $\pm$ 0.00013 & 1.66721 $\pm$ 0.00014 & 1.72170 $\pm$ 0.00016\\
$\Delta$ & -1.7 $\pm$ 2.2 \AA & 1.7 $\pm$ 2.3 \AA & 1.8 $\pm$ 1.8 \AA & 0.0 $\pm$ 2.0 \AA & 2.2 $\pm$ 2.1 \AA\\
\hline
\end{tabular}
\end{center}
\captionlistentry{}
\label{bandshifttab}
\end{table}

\begin{figure}[ht!]
\begin{center}
\includegraphics[scale=0.65,trim=0cm 0cm 0cm 0cm,clip=true]{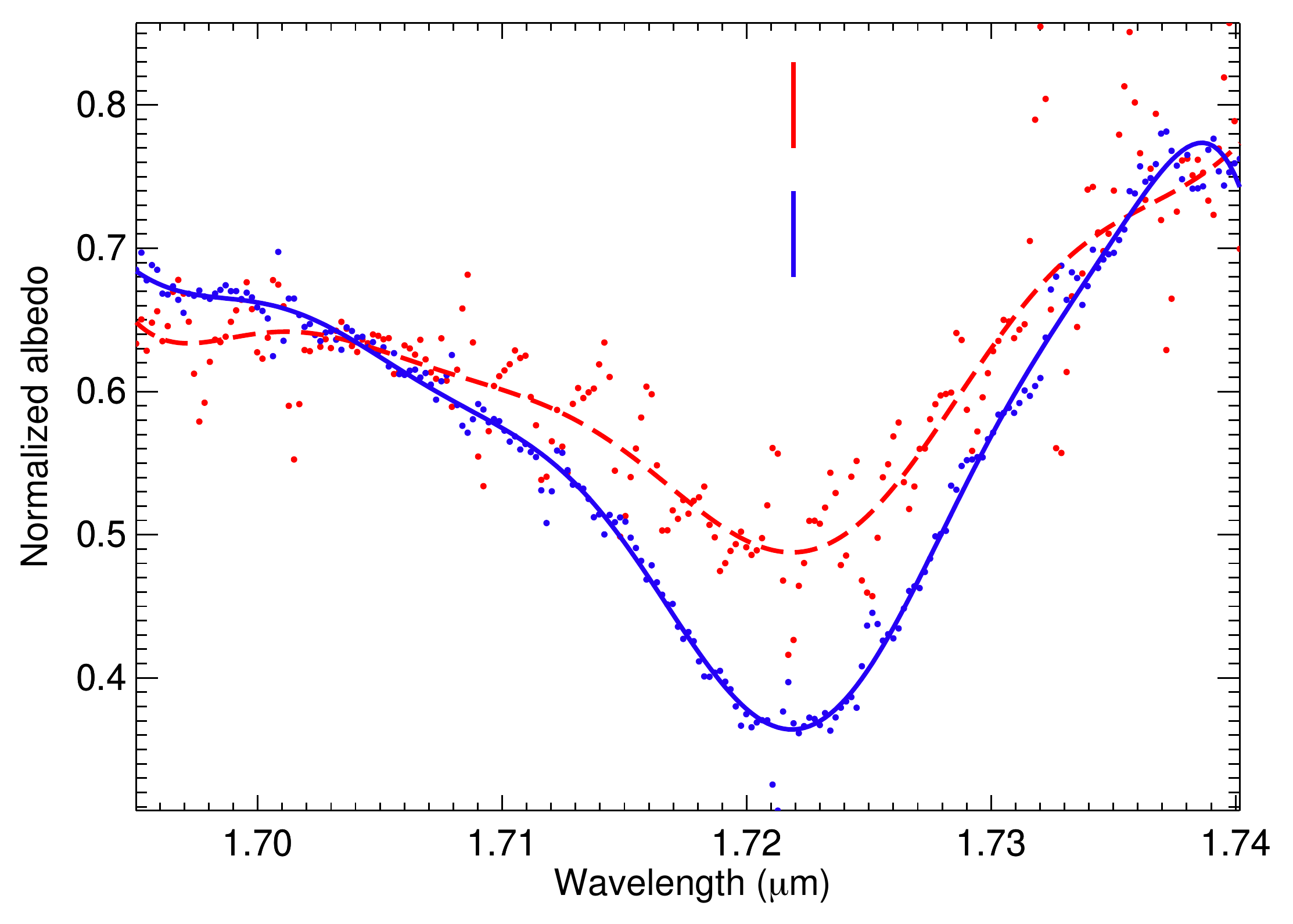}
\caption{Comparison of the Fourier fits for the 1.72 $\mu$m CH$_4$ band from the night-averaged spectra of 2014-06-17 (red dashed lined) and 2015-08-19 (blue solid line). This is the only matched pair with a statistically significant increase in the 1.72 $\mu$m band area (Figure~\ref{bandarea}). The positions of the band minima are marked by the vertical red and blue lines; both minima are at 1.72192 $\mu$m (Table~\ref{bandshifttab}). Using the correlation technique of Tegler et al. (2008), this particular pair of 1.72 $\mu$m bands presented a statistically significant blueshift over time ($>$5-$\sigma$), which is easily ruled out through visual inspection. \label{bandshift}}
\end{center}
\end{figure}

\item {\it Preferential sublimation of N$_2$}: The thermodynamically favored option, which agrees with the results of thermal and atmospheric models (Hansen and Paige, 1996; Young, 2013; Hansen et al., 2015; Bertrand and Forget, 2016), is the preferential sublimation of N$_2$ ice as northern hemisphere summer approaches. As mentioned previously, N$_2$ is more volatile than CH$_4$ (e.g., Fray and Schmitt, 2009) and will therefore sublimate prior to CH$_4$ as the surface temperature increases due to increased insolation. Protopapa et al. (2015) present laboratory-derived optical constants for different concentrations of N$_2$ and CH$_4$ in solution at varying temperatures. Their results indicate that for solutions dominated by CH$_4$ (CH$_4$:N$_2$), the band center shift as a function of CH$_4$ concentration follows a very shallow slope, with a steeper slope when CH$_4$ concentration decreases below 50\%; see Figure 11 of Protopapa et al. (2015). For the 2014-06-17/2015-08-19 matched pair, the measured band shift for the 1.72 $\mu$m CH$_4$ band was 0.0 \AA~with an uncertainty of 2.6 \AA; this uncertainty equates to a 9.5\% change in CH$_4$ concentration, using the shallower linear fit to CH$_4$-dominated mixtures (Protopapa et al., 2015). For the same uncertainty, a 5-$\sigma$ detection would require a band shift of 13.0 \AA, which corresponds to a 56.8\% change in CH$_4$ concentration, assuming the initial or final CH$_4$ concentration is 100\%.\\
\indent A band shift of 0 \AA~does not agree with the measured increase in the 1.72 $\mu$m band area between 2014-06-17 and 2015-08-19 and the assumption that N$_2$ sublimation is the cause. It is therefore worth examining what a redshift at the 1-$\sigma$ level (2.6 \AA, or a 9.5\% increase in CH$_4$ concentration) would produce in terms of band area changes. Brunetto et al. (2008) report full width at half-maximum (FWHM) values for 12 CH$_4$ features between 1.33 and 3.55 $\mu$m using different mixtures of CH$_4$ and N$_2$. Specifically, they report FWHMs for mixtures with CH$_4$ concentrations of 100, 91, 50, and 9\%. A change from 91\% to 100\% CH$_4$ concentration approximately matches a band shift at the 1-$\sigma$ level and the FWHMs increase roughly linearly with increasing CH$_4$ concentration. To determine the expected band area difference, we modeled the 1.72 $\mu$m band on each date of the matched pair as a Gaussian with standard deviations equal to the FWHMs divided by 2$\sqrt{2\mathrm{ln}2}$. The band center was set to 1.72 $\mu$m for both bands. Using the same wavelength range for the band area calculation (1.718-1.723 $\mu$m), we find a 6.5\% increase in band area from 2014 to 2015 (measured as the difference divided by the value in 2014). This is less than the 19\% increase reported for this particular matched pair (Table~\ref{bandareatab}). However, the FWHMs from Brunetto et al. (2008) were reported for an ice temperature of 16 K, much lower than the $\sim$40 K temperature expected for Pluto's surface. Protopapa et al. (2015) show that the CH$_4$ band FWHMs increase with increasing temperature. Additionally, higher CH$_4$ concentrations result in larger FWHMs. Assuming some combination of the above effects in broadening the observed 1.72 $\mu$m band, and applying a uniform broadening factor to the FWHM values in the Gaussian models, a 19\% change in band area is achievable. Due to the degeneracy in temperature and CH$_4$ concentration needed to reproduce the properties of the observed bands, we are unable to determine the initial and final values of CH$_4$ concentration between the components of the matched pair.\\
\indent Based on spectral data obtained with New Horizons in July 2015, different mixing states of N$_2$ and CH$_4$ appear to be latitudinally segregated on Pluto (Protopapa et al., 2017). Latitudes between 20 and 35$^{\circ}$N are dominated by CH$_4$:N$_2$, 35-55$^{\circ}$N is dominated by N$_2$:CH$_4$, and 55-90$^{\circ}$N is dominated by CH$_4$:N$_2$; see Figures 4 and 5 in Protopapa et al. (2017). The concentration of N$_2$ is probably lower between 55 and 90$^{\circ}$N compared to 20 to 35$^{\circ}$N, due to net loss of N$_2$ from northern latitudes towards southern latitudes (e.g., Hansen et al., 2015; Lewis et al., 2021). The increase in CH$_4$ absorption between 2014-06-17 and 2015-08-19, $\sim$1 month after the New Horizons flyby, suggests that the northern hemisphere supported higher N$_2$ concentrations only a year before the flyby. Older observations suggest the north polar region supported even larger quantities of N$_2$ in the past. Visible maps of Pluto from 1994 show that the northern polar region was highly reflective, with the interpretation that this was a deposit of N$_2$ ice (Stern et al., 1997).\\
\indent As described by Protopapa et al. (2017), the increase in CH$_4$ concentration is likely due to two different effects: (1) sublimation of N$_2$ from the CH$_4$:N$_2$-dominated northernmost latitudes and (2) sublimation of N$_2$ from the northern and southern edges of the N$_2$:CH$_4$-dominated collar. Both of these effects would have to occur in order to produce a change in CH$_4$ concentration larger than a few percent. This is because N$_2$ and CH$_4$ are only partially miscible in each other, with a maximum concentration of 5\% CH$_4$ in N$_2$:CH$_4$ and 3\% N$_2$ in CH$_4$:N$_2$ (Prokhvatilov and Yantsevich, 1983; Protopapa et al., 2015). The approximate northern boundary between the CH$_4$:N$_2$- and N$_2$:CH$_4$-dominated regions roughly corresponded to the southernmost latitude that was in constant sunlight starting in 2005. As the sub-solar latitude migrates northwards, the area in constant sunlight expands. Regions that have experienced constant sunlight longer have lost a larger quantity of N$_2$ ice and are therefore dominated by less-volatile CH$_4$. Latitudes just to the south of this boundary are in the process of transitioning to being CH$_4$:N$_2$-dominated, leading to the expansion of this region below 55$^{\circ}$N. Protopapa et al. (2017) suggest an additional, slower N$_2$ sublimation front moving northward across this collar, with the two fronts meeting within the next decade, resulting in the disappearance of the collar. This southern sublimation front may have been responsible for the initial creation of the collar, with N$_2$ sublimating from more equatorial latitudes and depositing in the north, but now increased insolation in the northern hemisphere is leading to the collar's steady destruction. Over time, prolonged periods of higher insolation should lead to the sublimation of CH$_4$ from the northern hemisphere and the generation of a ``polar bald spot" (Hansen and Paige, 1996; Hansen et al., 2015), but this was not detected during the 2014-2017 time frame.

\end{enumerate}

\begin{figure}[ht!]
\begin{center}
\includegraphics[scale=0.6,trim=0cm 0cm 0cm 0cm,clip=true]{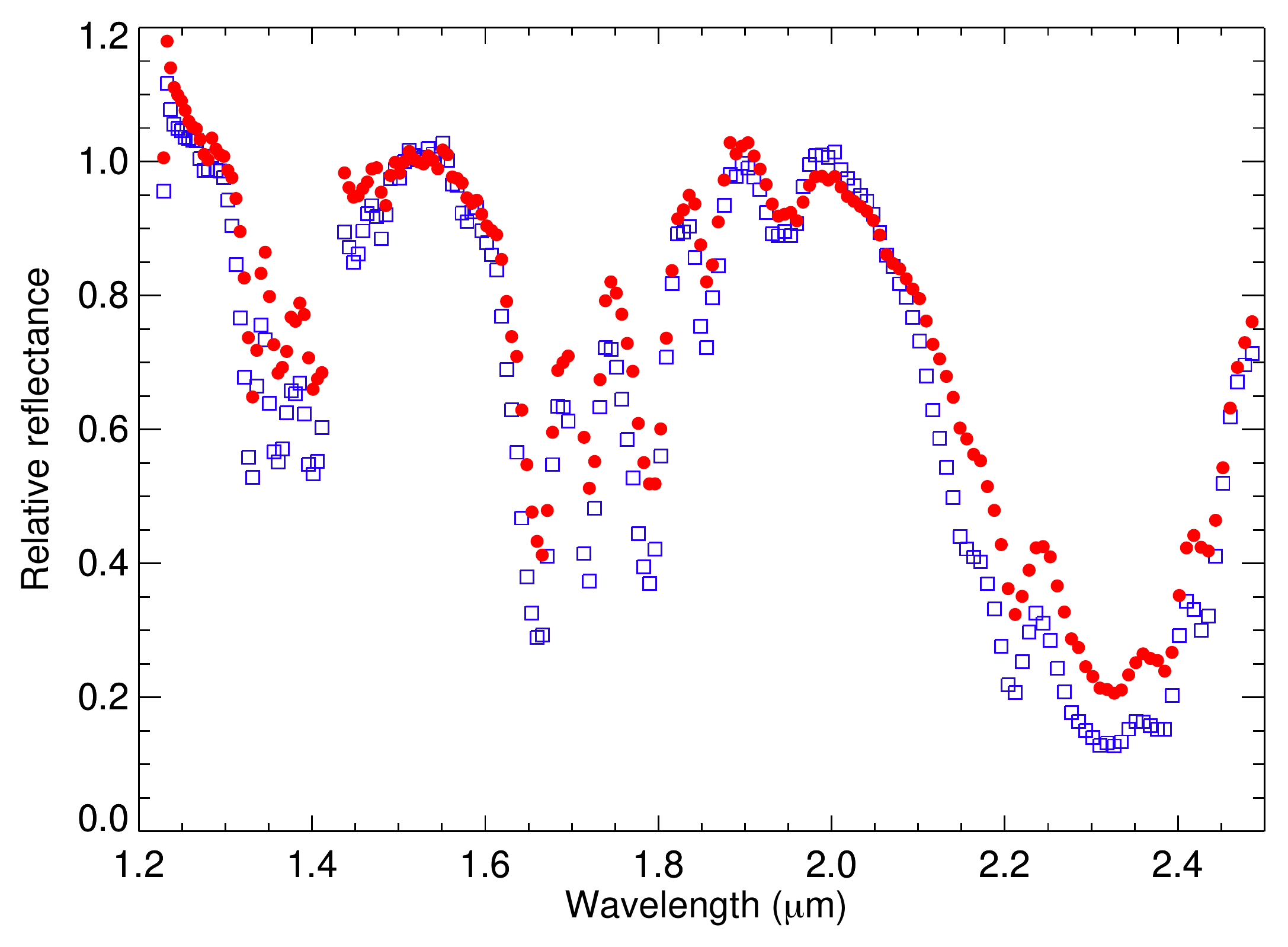}
\caption{Comparison of the integrated spectrum of Sputnik Planitia (blue squares) to the integrated spectrum of the remainder of Pluto's anti-Charon hemisphere (red circles). These spectra were extracted from New Horizons LEISA data, specifically the P\_LEISA\_Alice\_2a and P\_LEISA\_Alice\_2b scans obtained on July 14$^{\mathrm{th}}$, 2015. To extract spectral information only from pixels within Sputnik Planitia, we identified all pixels with altitudes lower than -1 km using the digital elevation model (DEM; e.g., Schenk et al., 2018) between 147 and 200$^{\circ}$ longitude. The total integrated flux across the anti-Charon hemisphere was calculated by summing all pixels with incidence and emission angles $<$80$^{\circ}$. The non-Sputnik spectrum was determined by subtracting the integrated Sputnik spectrum from the total spectrum. The two spectra were then normalized to a continuum region at 1.5 $\mu$m. Since the spectra are normalized, the stronger CH$_4$ absorption features in the Sputnik spectrum show the larger contribution Sputnik makes to these features on a per-area basis. Changes in surface composition on the non-Sputnik portion of Pluto are therefore more difficult to detect in disk-integrated spectra that include Sputnik.\label{sputnik}}
\end{center}
\end{figure}

It is important to note that the observed changes in Pluto's northern hemisphere are not confined to the sub-observer hemisphere centered at 280$^{\circ}$, simply that the changes are easier to detect without spectral contributions from Sputnik Planitia. Sputnik dominates the disk-integrated spectra of the majority of the matched pairs included in this work (Figure~\ref{sputnik}) and, while it does undergo seasonal changes, it evolves on much longer timescales (Bertrand et al., 2018). The sub-observer hemisphere that presents the only statistically significant increase in CH$_4$ band areas is centered at $\sim$280$^{\circ}$ and contains a small portion of Sputnik Planitia on the limb and a large portion of Cthulhu Macula (Figure~\ref{subobshemis}). The matched pair centered at $\sim$170$^{\circ}$, which covers approximately the same time span, does not contain any portion of Sputnik Planitia or Cthulhu Macula and did not present an increase in CH$_4$ at the $>$5-$\sigma$ level (though it was very close). It is possible that the lack of a more significant band area increase for this matched pair is due to re-deposition of N$_2$ at more southerly latitudes; unfortunately, this cannot be evaluated because New Horizons did not obtain spectral observations of the non-encounter (sub-Charon) hemisphere. Conversely, Cthulhu Macula is a low-albedo, higher-temperature region and therefore is not a site for volatile deposition. The unique combination of minimal contributions from Sputnik and the presence of Cthulhu likely led to the high-significance changes in CH$_4$ band areas observed between 2014-06-17 and 2015-08-19.}

\section{Summary}
\doublespacing{In this investigation we used near-infrared spectra of approximately the same sub-observer hemisphere on Pluto (matched pairs) to evaluate temporal changes on 1-, 2-, and 3-year timescales between 2014 and 2017. This resulted in 4 matched pairs and 1 matched quartet and the following results:
\begin{itemize}
\item A statistically significant ($>$5-$\sigma$) increase in the band areas of the 1.66 and 1.72 $\mu$m CH$_4$ absorption features between 2014-06-17 and 2015-08-19. The sub-observer hemisphere was centered on 280$^{\circ}$E. No other CH$_4$ features presented a $>$5-$\sigma$ increase between components of any of the other matched pairs.

\item CH$_4$ band center shifts were also computed for each of the absorption features considered in this investigation, with no statistically significant shifts detected.

\item A change in CH$_4$ grain size and deposition of CH$_4$ from the atmosphere were ruled out as causes for the increased CH$_4$ band areas. The preferred explanation is sublimation of N$_2$ from the northernmost latitudes on Pluto as northern summer solstice approaches, leading to an increase in CH$_4$ concentration.

\item Uncertainties on the band shift calculations are large enough that significant changes in CH$_4$ concentration could occur and not be detected using these data.

\item The sub-observer hemisphere for which the increase occurred was unique among all the matched pairs because it included only a small portion of Sputnik Planitia and a large portion of Cthulhu Macula. The majority of the other matched pairs were centered on Sputnik Planitia. This combination of surface features likely made it easier to detect surface composition changes in the north polar region on this particular sub-observer hemisphere.
\end{itemize}

These results indicate that Pluto is experiencing rapid seasonal changes, compared to its 248-year orbital period, in the lead up to northern hemisphere summer solstice in 2029. The loss of more-volatile N$_2$ from the northern hemisphere is potentially a harbinger of future loss of less volatile CH$_4$. Continued ground-based spectral monitoring of Pluto is necessary to detect the onset of CH$_4$ loss and to evaluate further seasonal changes, which could lead to advances in global circulation models, atmospheric studies, and thermal inertia determinations.
}

\section*{Acknowledgements}
\doublespacing{This paper made use of data obtained with the Apache Point Observatory (APO) 3.5-meter telescope, which is owned and operated by the Astrophysical Research Consortium (ARC). The authors would like to thank the APO telescope operators who helped keep this observing program running smoothly: Alaina Bradley, Jack Dembicky, Candace Gray, Russet McMillan, and Theodore Rudyk. The authors would also like to thank Fran Bagenal, John Bally, and Dave Brain at the University of Colorado for their support of this student-led observing program, and Mike Skrutskie and Jhett Bordwell for their advice during the data reduction and analysis. The authors would also like to recognize the feedback and suggestions of two anonymous reviewers. B. Holler would like to acknowledge funding from NASA NESSF 14-PLANET14F-0045 and NASA SSO 16-SSO16\_2-0030. S. Protopapa would like to acknowledge funding from NASA SSW 80NSSC19K0554.}

\section*{References}
\begin{hangparas}{0.25in}{1}
\doublespacing{Bertrand, T., Forget, F., 2016. Observed glacier and volatile distribution on Pluto from atmosphere-topography processes. Nature 540, 86-89.

Bertrand, T., et al., 2018. The nitrogen cycles on Pluto over seasonal and astronomical timescales. Icarus 309, 277-296.

Binzel, R. P., et al., 2017. Climate zones on Pluto and Charon. Icarus 287, 30-36.

Brunetto, R., et al., 2008. Integrated near-infrared band strengths of solid CH$_4$ and its mixtures with N$_2$. ApJ 686, 1480-1485.

Buratti, B. J., et al., 2021. Pluto in glory: Discovery of its huge opposition surge. GRL 48, e92562.

Cruikshank, D. P., et al., 1976. Pluto: Evidence for methane frost. Science 194, 835-837.

Cushing, M. C., et al., 2004. Spextool: A spectral extraction package for SpeX, a 0.8-5.5 micron cross-dispersed spectrograph. PASP 116, 362-376.

Eluszkiewicz, J., 1991. On the microphyscial state of the surface of Triton. JGR 96, 19217-19229.

Fray, N., Schmitt, B., et al., 2009. Sublimation of ices of astrophysical interest: A bibliographic review. P\&SS 57, 2053-2080.

Gladstone, G. R., et al., 2016. The atmosphere of Pluto as observed by New Horizons. Science 351, aad8866.

Gradie, J. C., et al., 1980. The effects of photometric geometry on spectral reflectance. Lunar and Planetary Science Conference XI, 357-359.

Grundy, W. M., Buie, M. W., 2001. Distribution and evolution of CH$_4$, N$_2$, and CO ices on Pluto's surface: 1995 to 1998. Icarus 153, 248-263.

Grundy, W. M., et al., 2002. The temperature-dependent spectrum of methane ice I between 0.7 and 5 $\mu$m and opportunities for near-infrared remote thermometry. Icarus 155, 486-496.

Grundy, W. M., et al., 2013. Near-infrared spectral monitoring of Pluto's ices: Spatial distribution and secular evolution. Icarus 223, 710-721.

Grundy, W. M., et al., 2014. Near-infrared spectral monitoring of Pluto's ices II: Recent decline of CO and N$_2$ ice absorptions. Icarus 235, 220-224.

Grundy, W. M., et al., 2016. Surface compositions across Pluto and Charon. Science 351, aad9189.

Hansen, C. J., Paige, D. A., 1996. Seasonal nitrogen cycles on Pluto. Icarus 120, 247-265.

Hansen, C. J., et al., 2015. Pluto's climate modeled with new observational constraints. Icarus 246, 183-191.

Hapke, B., 2012. Theory of Reflectance and Emittance Spectroscopy, second ed. Cambridge University Press.

Hofgartner, J. D., et al., 2018. A search for temporal changes on Pluto and Charon. Icarus 302, 273-284.

Lewis B. L., et al., 2021. Distribution and energy balance of Pluto's nitrogen ice, as seen by New Horizons in 2015. Icarus 356, 113633.

Lorenzi, V., et al., 2016. The spectrum of Pluto, 0.40-0.93 $\mu$m. I. Secular and longitudinal distribution of ices and complex organics. A\&A 585, A131.

Meza, E., et al., 2019. Lower atmosphere and pressure evolution of Pluto from ground-based stellar occultations, 1988-2016. A\&A 625, A42.

Molaro, J. L., et al., 2018. Microstructural evolution of solar system ices through sintering. 49$^{\mathrm{th}}$ Lunar and Planetary Science Conference, 2977.

Olkin, C.B., et al., 2021. Colors and photometric properties of Pluto. In: Stern, S. A., Moore, J. M., Grundy, W. M., Young, L. A., Binzel, R. P. (Eds.), The Pluto System After New Horizons. University of Arizona Press, Tucson, pp. 147-164.

Pitman, K. M., et al., 2017. Coherent backscattering effect in spectra of icy satellites and its modeling using multi-sphere T-matrix (MSTM) code for layers of particles. P\&SS 149, 23-31.

Prokhvatilov, A. I., Yantsevich, L. D., 1983. X-ray investigation of the equilibrium phase diagram of CH$_4$-N$_2$ solid mixtures. Sov. J. Low Temp. Phys. 9, 94-98.

Protopapa, S., et al., 2015. Absorption coefficients of the methane-nitrogen binary ice system: Implications for Pluto. Icarus 253, 179-188.

Protopapa, S., et al., 2017. Pluto's global surface composition through pixel-by-pixel Hapke modeling of New Horizons Ralph/LEISA data. Icarus 287, 218-228.

Quirico, E., Schmitt, B., 1997. Near-infrared spectroscopy of simple hydrocarbons and carbon oxides diluted in solid N$_2$ and as pure ices: Implications for Triton and Pluto. Icarus 127, 354-378.

Schenk, P. M., et al., 2018. Basins, fractures and volcanoes: Global cartography and topography of Pluto from New Horizons. Icarus 314, 400-433.

Stern, S. A., Trafton, L. Constraints on bulk composition, seasonal variation, and global dynamics of Pluto's atmosphere. Icarus 57, 231-240.

Stern, S. A., et al., 1997. HST high-resolution images and maps of Pluto. AJ 113, 827-843.

Stern, S. A., et al., 2015. The Pluto system: Initial results from its exploration by New Horizons. Science 350, aad1815.

Tegler, S. C., et al., 2008. Evidence of N$_2$-ice on the surface of the icy dwarf planet 136472 (2005 FY9). Icarus 195, 844-850.

Verbiscer, A., et al., 2020. A tale of two sides: Pluto's opposition surge in 2018 and 2019. 14$^{th}$ Europlanet Science Congress, EPSC2020-546.

Wilson, J. C., et al., 2004. Mass producing an efficient NIR spectrograph. Proc. SPIE 5492, 1295-1305.

Young, L. A., 2013. Pluto's seasons: New predictions for New Horizons. ApJL 766, L22.

Zangari, A., 2015. A meta-analysis of coordinate systems and bibliography of their use on Pluto from Charon's discovery to the present day. Icarus 246, 93-145.

}
\end{hangparas}

\end{document}